\documentclass[conference]{IEEEtran}
\IEEEoverridecommandlockouts
\usepackage{cite}
\usepackage[hyphens]{url}
\usepackage{hyperref}
\usepackage{amsmath,amssymb,amsfonts}
\usepackage{algorithmic}
\usepackage{graphicx}
\usepackage{textcomp}
\usepackage{xcolor}

\def\BibTeX{{\rm B\kern-.05em{\sc i\kern-.025em b}\kern-.08em
    T\kern-.1667em\lower.7ex\hbox{E}\kern-.125emX}}
\begin{document}

\title{Deception for Cyber Defence:\\ Challenges and Opportunities }

\makeatletter
\newcommand{\linebreakand}{%
  \end{@IEEEauthorhalign}
  \hfill\mbox{}\par
  \mbox{}\hfill\begin{@IEEEauthorhalign}
}
\makeatother
\author{
    \IEEEauthorblockN{David Liebowitz\IEEEauthorrefmark{3}\IEEEauthorrefmark{4},
    Surya Nepal\IEEEauthorrefmark{1}\IEEEauthorrefmark{2},
    Kristen Moore\IEEEauthorrefmark{1}\IEEEauthorrefmark{2},
    Cody J. Christopher\IEEEauthorrefmark{1}\IEEEauthorrefmark{2},
    Salil S. Kanhere \IEEEauthorrefmark{4}}
    \linebreakand
    \IEEEauthorblockN{David Nguyen\IEEEauthorrefmark{1}\IEEEauthorrefmark{2}\IEEEauthorrefmark{4},
    Roelien C. Timmer\IEEEauthorrefmark{1}\IEEEauthorrefmark{2}\IEEEauthorrefmark{4},
    Michael Longland\IEEEauthorrefmark{1}\IEEEauthorrefmark{2}\IEEEauthorrefmark{4},
    Keerth Rathakumar\IEEEauthorrefmark{1}\IEEEauthorrefmark{2}\IEEEauthorrefmark{4}}
    \linebreakand
    \IEEEauthorblockA{
    \IEEEauthorrefmark{1}Data61, CSIRO\\
        Australia\\
        \{first.last\}@data61.csiro.au
    }
    \and
    \IEEEauthorblockA{
        \IEEEauthorrefmark{2}Cyber Security \\
        Cooperative Research\\
         Centre Australia\\
    }
    \and
    \IEEEauthorblockA{
    \IEEEauthorrefmark{3}Penten Pty Ltd\\
        Canberra, Australia\\
        \{first.last\}@penten.com
    }
    \and
    \IEEEauthorblockA{
        \IEEEauthorrefmark{4}UNSW\\
        Sydney, Australia\\
        \{first.last\}@unsw.edu.au
    }
}

\maketitle

\begin{abstract}
Deception is rapidly growing as an important tool for cyber defence, complementing existing perimeter security measures to rapidly detect breaches and data theft. 
One of the factors limiting the use of deception has been the cost of generating realistic artefacts by hand. Recent advances in Machine Learning have, however, created opportunities for scalable, automated generation of realistic deceptions. This vision paper describes the opportunities and challenges involved in developing models to mimic many common elements of the IT stack 
for deception effects.
\end{abstract}


\section{Introduction}

The digital revolution has touched every aspect of our lives. Precision health, digital agriculture, autonomous vehicles, digital governments and services are a few examples. The COVID-19 pandemic taught us that we could rely on the internet-enabled digital world to do many things that we did not think possible even a year ago.  But as the old saying goes, there is no free lunch – and this is especially true in the digital world. There are many challenges, but  cybersecurity stands front and centre as one of the most significant. Cybersecurity risks pose a threat to the continued growth and success of the internet-enabled digital world. 

Governments, academia and industry have recognised this challenge. An enormous amount of money has been spent on cybersecurity, with forecasts of expenditure exceeding USD 150 billion worldwide in 2021\cite{gartner_spending_2021}. Despite these efforts, breaches and data theft continue. Existing cyber defence solutions are like a tailor's patchwork. Most of them are reactive and are not able to deal with sophisticated attacks. The Target Corporation breach of 2013 exemplifies the problem of existing reactive solutions\cite{weiss2015target}.  

Target was compromised by a sophisticated attack that resulted in a period of approximately a month with intruders active on their network. This included access to their point-of-sale system over the busy Black Friday shopping period\cite{targetpressrelease2013}. The breach resulted in the theft of 40 million credit card numbers and 70 million personal records, and an estimated card reissue cost of USD 200 million\cite{krebs2014target}. Lawsuits have dragged on for years \cite{reuters2017target} and the event is still used as a cautionary case study \cite{harrell2017synergistic, plachkinova2018security}. 

The Target case study gives us two important insights into reactive cyber defence solutions. First, Target's security system raised alarms at the initial breach, but these were reportedly never investigated\cite{schwartz2014target}. The most likely reason for this is simple alarm fatigue. The patchwork-based reactive solutions generate a massive number of alerts. It is impossible to investigate all of them, and the prioritisation of alerts remains a challenge. Second, the damage was compounded by the dwell time. Prolonged, undetected access permitted movement across the network and allowed the sustained harvesting of information. According to the 2021 Cost of a Data Breach Report \cite{ponemon2021}, the {\em average} time taken to identify a breach has risen to 212 days. We have witnessed several real-world case studies in the last decade with similar attributes and results\cite{weiss2015target}. The time is upon us to rethink our approaches to cybersecurity and incorporate more proactive cyber defence tools to our existing security arsenal. 

In this vision article, we argue that it is time to revisit cyber deception. We believe that this is one of the most potent defensive security tools at our disposal and is arguably one of the most underutilised.  Deception is an attempt to manipulate the beliefs of others in order to influence their behaviour. In the cyber domain, this usually means creating and deploying a {\em honeypot} of some sort - a fake digital resource that mimics some characteristics of real resources~\cite{spitzner2003-1-honeypots}.  In doing so, the cyber defender tricks intruders, data thieves or malicious insiders into behaviour that reveals their presence and possibly more information.

Take, for example, a honeypot appliance on a network, mimicking a server \cite{fraunholz2018defending}. It is possible that a systems administrator {\em might} accidentally try to access it during routine maintenance. However, an access attempt is far more likely due to intruder lateral movement or attempted data theft.  This ability to highlight unauthorised activity makes deception a valuable breach discovery mechanism. Its intrinsically low false positive alarm rates can mitigate some of the effects of the alert fatigue problem, which helps solve the breach discovery problem, which in turn solves the problem of dwell time.

In addition to its role in counteracting alert fatigue and reducing dwell time, deceptions can become a mechanism to drive adversary interactions that can provide intelligence on their intent, or their tactics, tools and procedures (TTPs). While simpler honeypots can be very effective for breach and theft detection, they do not drive the level of interaction necessary for these additional effects. Simple honeypots can also become familiar to adversaries, who can begin to detect them without raising alarms. 

% discover the intent of intruders by having them engage with the honeypots. In particular, we can discover their intent by inference from their choice of which honeypot to engage with. Want the deception to be good enough to tell what they’re interested in by which honeypot they engage with. Or the environment is so real they deploy their TTPs. 

The key to engaging intruders to interact with deceptions is {\em realism}. 
Deceptive artefacts must be 
indistinguishable from the real subjects of mimicry, at least up to a level appropriate to the context and interaction. Generating such realistic honeypots by hand requires a significant amount of tedious work, which has impeded development of deceptive technologies until now.  
Indeed, we believe that the main challenge is to create deceptions automatically, so that they can be generated and deployed at the scale necessary to be 
useful as a defensive tool.
Recent advances in Machine Learning means it is now possible to model and create artifacts to simulate characteristics of many IT assets and processes. This vision paper presents our work-in-progress on such modeling.

We discuss some of the background to cyber deception in Section II.  Section III summarises some challenges and opportunities in using ML. It also highlights and provides a brief overview of our ongoing research and outcomes in this area. Finally, we present some discussions and concluding remarks. 

\section{Background}

{\bf What is Deception?} Deception is an attempt to manipulate the beliefs of others to influence their behaviour. The practice of deception is as old as life itself. The fierce competition to survive has produced breathtaking examples in the natural world ~\cite{stevens2016cheats}, such as camouflage predators like cuttlefish that can change the colour and texture of their skin to match their surroundings, insects indistinguishable from twigs when stationary, eye spots on moths that imitate much larger animals and plants that mimic pheromones or food scents to attract pollinating insects.

{\bf Deception in the Physical World:} Deception has also long been a key strategy in human conflict and warfare. Operation Bodyguard~\cite{smith2014overlord}, to take one famous example, was a World War II Allied deception plan implemented to mislead the German high command about the site of the D-Day landings in Normandy. The plan was named for Churchill’s observation that ”In wartime, truth is so precious that she should always be attended by a bodyguard of lies.” Large scale, coordinated activity with dummy vehicles and aircraft, fake radio signals and the use of double agents created the illusion of Allied invasion forces poised to attack various parts of the European coastline. It successfully concealed the real landing site and prompted diversion of defensive forces away from it, making a substantial contribution to Allied victory in the Battle of Normandy and the subsequent end of the war. We can protect information resources with the very same concepts and techniques found in nature and expressed in conflict ~\cite{bell2003toward,whaley1982toward}, but adapted to the cyber domain.

{\bf Deception in the Cyber Domain:} The power of deception as a cyber defensive tool was demonstrated in the 1980s by Clifford Stoll. Stoll, an astronomer working as a systems administrator in the Lawrence Berkeley National Laboratory in California, was asked to investigate a 75 cent discrepancy in the mainframe accounting system. His examination revealed that the accounting anomaly was a trace left by an intruder who had broken into the system, and was also using it as a platform to compromise other systems, including some in the military. Stoll started logging every move made by the intruder, eventually involving a number of US and international law enforcement agencies in an effort to trace and apprehend him. They discovered that he was in Germany,
but had difficulty finding an exact location using the telephone call tracing technology of the time. In order to keep him online long enough, Stoll began creating documents
related to a fictional ”Strategic Defense Initiative”, writing a whole suite of convincing documents and letters. The ploy was a success, and led to the arrest of Markus Hess, a West
German citizen who broke into computer systems and sold the information to the KGB. Stoll published a paper describing his experiences ~\cite{stoll1988stalking} and expanded it in an engaging and accessible book, ”The Cuckoo’s Egg” ~\cite{stoll2005cuckoo}, that is highly recommended reading for anyone with an interest in cyber deception.

Research into deception techniques and implementation of honeypots has grown, led by organisations such as the Honeynet Project\footnote{https://www.honeynet.org}.  
Deception is also getting commercial attention now, with 
a number of companies gaining traction with deception products, including 
Attivo\footnote{https://www.attivonetworks.com}, Countercraft\footnote{https://www.countercraftsec.com}, Thinkst\footnote{https://thinkst.com} and TrapX\footnote{https://www.trapx.com}.  
Australian cyber company Penten is one of the participants in the work described in this paper. Penten\footnote{https://www.penten.com} has a deception based data theft detection offering called HoneyTrace\footnote{https://honeytrace.io} that allows clients to create customised fake documents, database entries, credit
card numbers, URLs and email addresses. The documents can trigger a beacon when opened,
and the URLs and email addresses are monitored for activity. Additionally, active search of social media and the dark web can detect the appearance of any of the artefacts, providing 
evidence of a breach. 

Creating realistic deceptive IT artefacts automatically and at scale remains difficult. In the next section, we describe some key challenges and the research opportunities offered by them.

\section{Challenges and Opportunities}
 The development of cyber deception technology involves creating the deceptive environment, as well as a model of the attacker and ways to measure their perception of the deception. Our approach is to develop rich and believable deceptive environments, populated with deceptive versions of assets that appear in the IT stack. In addition to creating these artefacts, we also simulate the behaviour of people and their interactions within the system. 
 
Since documents are often sought after by attackers, a particular focus within our group has been on the generation of documents and the simulation of people's interactions with them. The concept of a document honeyfile for intrusion detection was developed by Yuille~\cite{yuill2004honeyfiles}. A series of papers 
from the Intrusion Detection Systems Lab at Columbia further explored honeyfile creation and 
deployment, including some studies of honeyfile
characteristics~\cite{voris2012lost, voris2013bait, voris2015fox,bowen2009baiting,salem2011decoy}, and Whitham \cite{whitham2017automating} proposed automated topical content 
creation. 
Building on this work, our interest in documents has broadened 
over time 
from typical office documents to include a range of media that can be found on computing systems, including source code and web/wiki resources. We have also considered the file systems in which generated documents can be found.

Databases are also frequently the target of network intruders, as they are a likely source of high value data such as passwords, payment details and other sensitive personal information. Data breaches of this nature are among the most publicised as they constitute massive privacy breaches and put both individuals and organisations at a variety of risks.
There has also been research on creating fake data to populate database honeypots using 
approaches like rule mining~\cite{bercovitch2011honeygen} and deep learning with differential privacy~\cite{abay2019using}.

The task of automating the generation of these artefacts has naturally led us to focus primarily on three ML technologies for developing deceptive environments: language models (for creating fake textual content), temporal point processes (for simulating interaction events) and graph neural networks.

The recent development of the Transformer architecture and language
models like GPT-2~\cite{radford2018improving} have radically changed text generation. The new models have enabled realistic text synthesis with freely available pre-trained models. We use both  GPT-2 and earlier Recurrent Neural Network (RNN) approaches to sequence generation\cite{graves2013generating} to create document content. The temporal characteristics of behaviour are modelled using Temporal Point Processes (TPPs), where a number of neural approaches have improved the flexibility and application domain~\cite{shchur2021review}. Artefacts such as file systems and database schema can be modelled as graphs, so we use graph approaches like Graph Recurrent Attention Networks (GRAN)\cite{liao2019efficient} to develop bespoke models.

\subsection{Creating documents} 
Real world documents are not just text, but can include images, plots and hyperlinks. In addition, our notion of documents includes user data artefacts such as usage history and user files, system files that support the illusion of a live environment, graphics, and other proprietary artefacts like code repositories. 

\paragraph{Source code}
In \cite{nguyen2021honeycode}, we address the generation of honeycode - fake software repositories that 
look like real code when observed through a repository search engine or by command line file content display. In other words, the honeycode need only look like real code when briefly inspected, but does not need to compile. It does, however, have to include a realistic folder structure, file names and mix of 
file content. 

Modeling general source code is a challenge because of the rich distribution of languages found within a population of software repositories. The majority of repositories contain a mixture of languages that are either programming or natural (e.g. readme files).  Furthermore, these two forms of language can be found in the same file with inline code comments. Modelling programming languages is difficult due to its highly structured nature that requires complex and long range dependencies across blocks of code. We must also consider programming language dependencies across files and hierarchies in the repository structure.

\begin{figure}[tbp]
\centerline{\includegraphics[width=\linewidth]{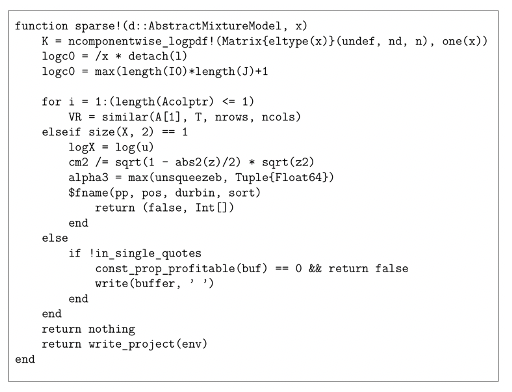}}
\caption{A code snippet generated by HoneyCode's file content generator. Several patterns emerge that create the illusion of realism including indentations, variable declaration, for-loops and type defining.
\label{fig:juliasnippet}}
\end{figure}

We regard a software repository as a structured amalgamation of three core components: directory structure, filenames and file content. GRAN was specialised to generate trees for modelling file systems, and conditional language generation used to improve the dependencies between different components. For example, we inject filename extensions (e.g. \verb|py|, \verb|txt|) into the file content generator to create consistency between the file types and their respective content. Recurrent Neural Networks trained on 3254 publicly available Julia software repositories from Github are used to generate filenames and file content. A sample of the generated code is provided in Figure~\ref{fig:juliasnippet} along with an overview of the three stage system in Figure~\ref{fig:softwarearch}.

Whilst this work is promising, the file content generator struggles with medium-long range dependencies due to the limited model capacity. This would certainly be improved using large scale Transformers~\cite{vaswani2017attention} for file content generation due to their ability to model distant dependencies through the use of a global receptive field, at the cost of requiring more training data. 

\begin{figure}[tbp]
\centerline{\includegraphics[width=\linewidth]{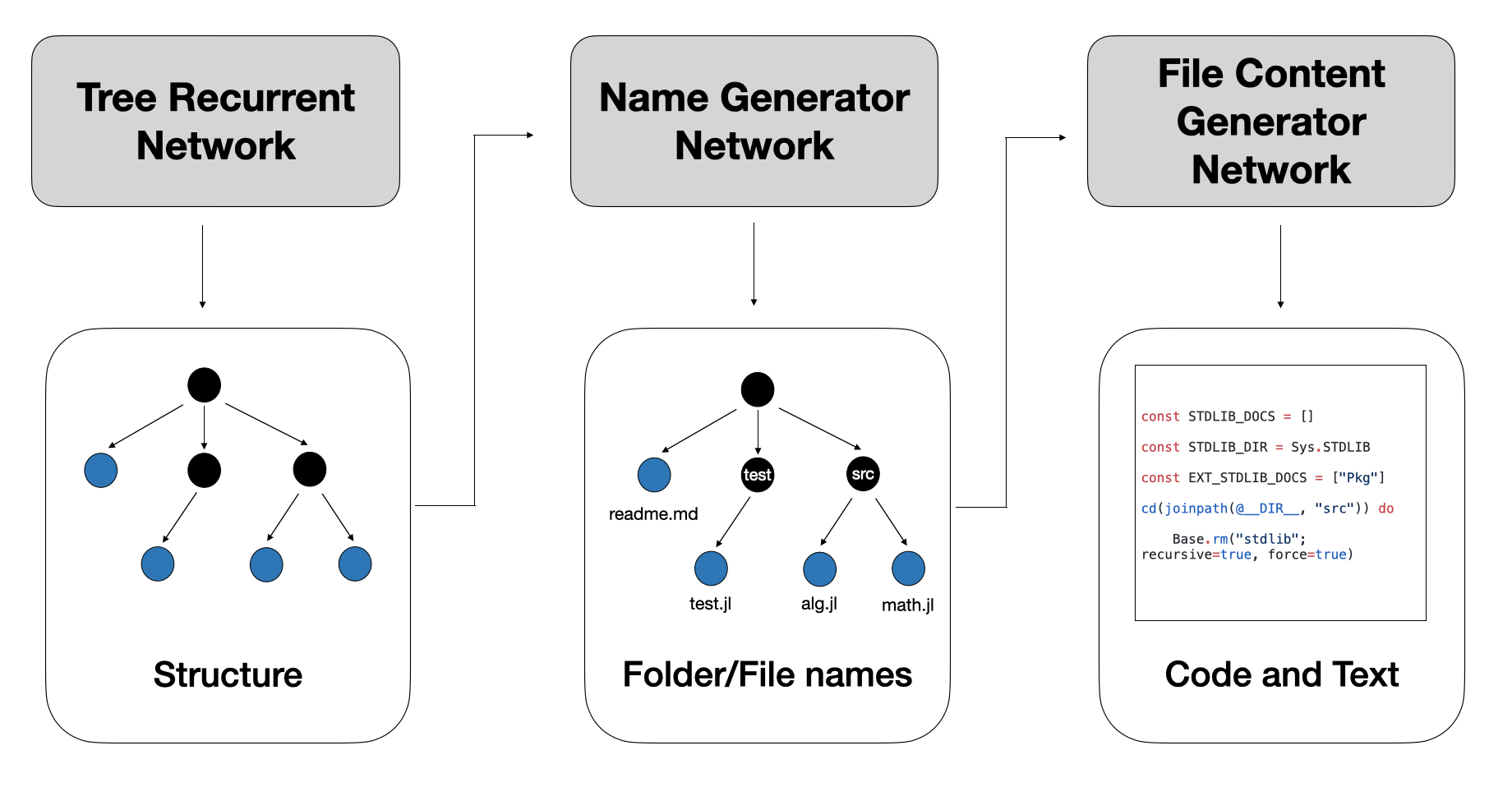}}
\caption{HoneyCode is composed of three different networks generating three core component, each using a different underlying structure to improve model inductive bias and realism. \label{fig:softwarearch}}
\end{figure}

\paragraph{Websites/Wikis}
Our work in \cite{longland2020wikigen} explores the use of GPT-2 to generate structured text in wiki-style web formats. Since these documents are written in Markdown, the document structure is encoded with the text. Section headings and other document structural elements are represented by Markdown symbols, which are used to render the documents. Fine-tuning GPT-2 with the Markdown symbols included in the vocabulary successfully enables 
generation of appropriately structured wiki-style pages. Synthetic articles will consistently begin with an introductory section, followed by a varying number of body sections and subsections, before concluding with references. Occasionally realism-breaking artefacts do emerge in synthetic articles, for example consecutive sections with the same name. Generated pages must be linked to create a realistic wiki, so a GRAN model is successfully trained to generate a page link network. 
The articles are then represented by the Smoothed Inverse Frequency (SIF) embedding~\cite{arora2017asimple} of their content and mapped to nodes in this page link network using a greedy heuristic algorithm. Articles with higher semantic similarity will then be more likely to share an edge in the link network.

The wiki generation process is shown in Figure~\ref{fig:wiki_arch}.
\begin{figure}
    \centering
    \includegraphics[width=.6\linewidth]{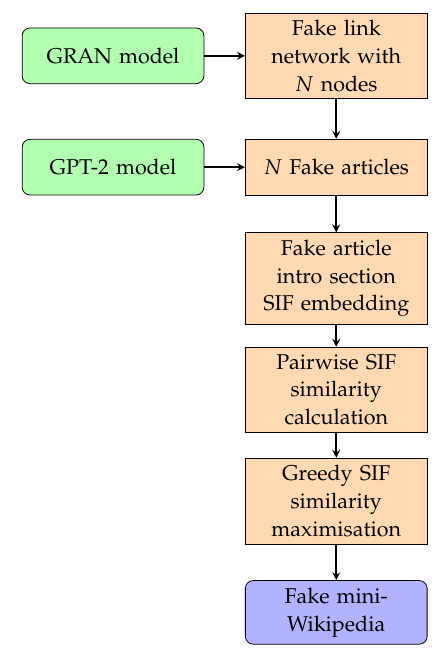}
    \caption{Wikigen is composed of two different models generating the article link network (GRAN) and the article content (GPT-2), followed by an optimisation procedure to have semantically similar articles be more likely to link to each other.}
    \label{fig:wiki_arch}
\end{figure}

\paragraph{Document layout} Graphical layouts are ubiquitous across a wide variety of assets found in corporate environments such as documents, posters and presentations. Instead of using human curated templates, recent research \cite{li2019layoutgan,jyothi2019layoutvae,lee2020neural} has focused on training deep learning models to generate realistic and authentic layout designs for the purpose of improving graphical editor recommendations. These learning-based approaches are promising for constructing more adaptive and dynamic honeyfiles.

 The earliest proposed layout generation approaches were based on variational auto-encoders (VAE) \cite{jyothi2019layoutvae,lee2020neural} or generative adversarial networks (GAN) \cite{li2019layoutgan} frameworks. Experiments indicate that these architectures are frequently subject to posterior collapse or instability in the latent space \cite{kikuchi2021constrained} leading to higher levels of sample degradation. In addition, these methods struggle to exhibit high level of output diversity due to their reliance on uni-modal prediction for object placement.  
 
 \begin{figure}[tbp]
\centerline{\includegraphics[width=\linewidth]{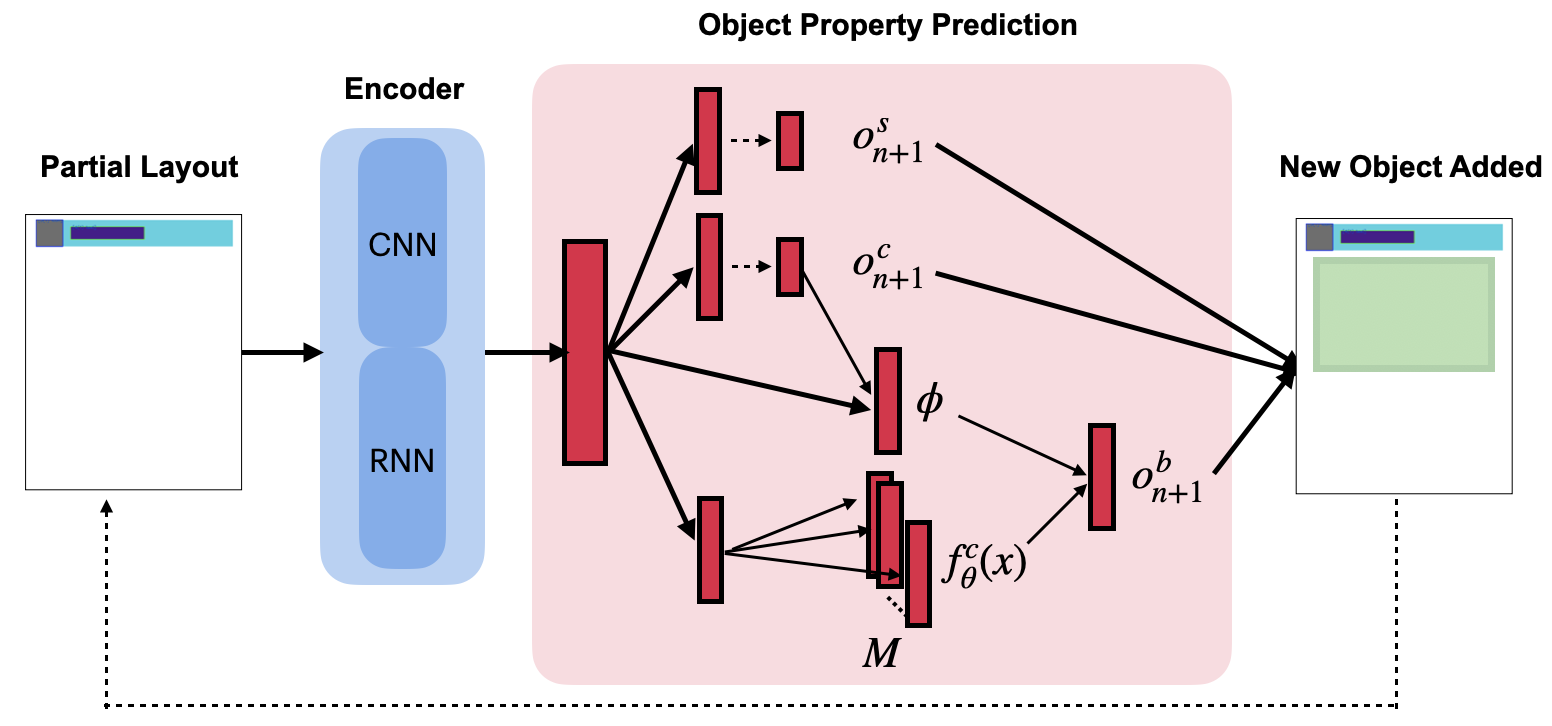}}
\caption{LayoutMCL is an autoregressive architecture that combines a multimodal encoder and triple head decoder to generate realistic and diverse layouts}
\label{fig:layoutmcl}
\end{figure}

Our recent work on LayoutMCL~\cite{nguyen2021diverse} demonstrates the feasibility of combining an auto regressive approach with multi choice learning~\cite{guzman2012multiple}. This unique architecture, shown in Figure~\ref{fig:layoutmcl}, generates significantly more stable, realistic and diverse layouts in comparison to its counterparts.  
This approach is also amenable to incorporating constraints such as a corporate logo or other consistent elements between layouts.  
These properties make LayoutMCL more suitable for fully automated honeyfile generation. 

\paragraph{Measuring the enticement of honeyfiles}
Looking beyond generation, it is useful to characterise honeypots to guide 
their creation and deployment. This is particularly true of honeyfiles, where there is a
great deal of flexibility in the content, appearance and placement of the deceptions. 
%There are, as Spitzner pointed out in the insider threat %context\cite{spitzner2003-1-honeypots}, two problems to solve to implement a deception 
%strategy with a honeypot: redirection to the honeypot and having a realistic honeypot to %interact with.
A number of honeyfile characteristics have been considered~\cite{voris2013bait, whitham2017automating}, including enticement (or enticingness), conspicuousness, believability and realism. We are primarily interested in two metrics: enticement as a measure of how well a honeyfile can attract the attention of an intruder, and realism as a measure of how plausible a mimic it is.

It is necessary to construct a model of the adversary in order to meaningfully develop
such metrics, and make assumptions about how they will find and interact with the honeyfile. 
If we assume that an intruder is trying to steal documents from an organisational document repository, we can design honeyfile text with the repository search interface in mind. Honeyfile content, to be enticing, should mimic the topics appearing in the real documents they are protecting so that the honeyfiles will be encountered by someone searching for those topics. 
This also means accounting for the fact that search engines often show snippets of text containing the search terms in the results, so the fake text should appear realistic and at least locally coherent. 
Earlier work on enticement~\cite{whitham2017automating} measured common word counts as a 
basis for a metric. Since this approach does not account for paraphrasing,  
we have introduced a measure called the Topic Semantic Matching (TSM) enticement score~\cite{timmer2022tsm}. TSM extracts the main topics of a repository, also know as local context, using topic modelling. Next, we calculate the similarity between all the extracted topics and the words in the deception file. This similarity uses a
 vector space representation of the words in which semantically similar words are close together, so it can detect similarity even when different words with similar meanings
 are used. 
A visualisation of the concept of TSM is shown in Figure~\ref{fig:tsm_visualisation}. Experiments show that the TSM measure achieves a higher score when a deception file of a specific theme is compared to a local context of the same theme as shown in Figure~\ref{fig:tsm_distribution}. 

\begin{figure}[tbp]
\centerline{\includegraphics[width=\linewidth]{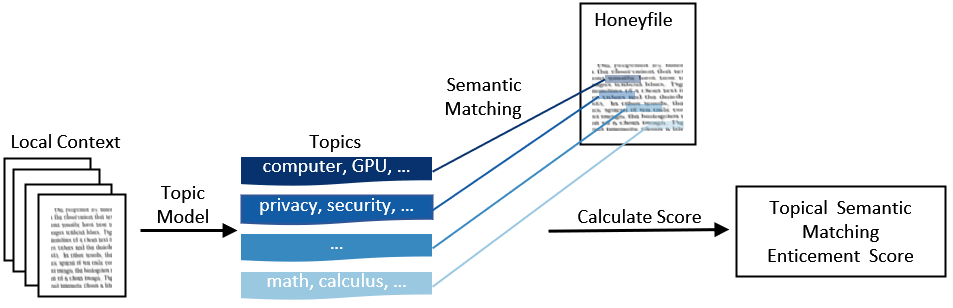}}
\caption{Visualisation of the Topic Semantic Matching enticement score for deception files.\label{fig:tsm_visualisation}}
\end{figure}

\begin{figure*}[tbh]
\centerline{\includegraphics[width=\textwidth]{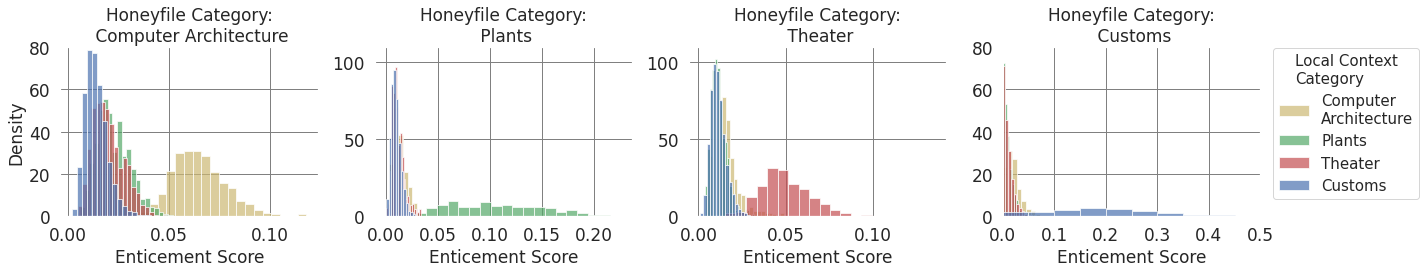}}
\caption{The distribution of the Topic Semantic Matching enticement score.}
\label{fig:tsm_distribution}
\end{figure*}

We are also interested in the presence of sensitive information, potentially derived from 
training or fine-tuning a language model on real documents, appearing in a honeyfile. This problem has been studied in a more general setting associated with classifying document 
sensitivity automatically. Typical models are based on the word occurrence and corresponding association rules~\cite{cho2008detectingprivacy, sanchez2016c}, and more recent research~\cite{neerbek2018detecting} experimented with RNNs to detect sensitive sentences.

Recent use of fine-tuning on pre-trained Transformer language models has proven effective on a
range of NLP tasks, and we have subsequently tested this idea on sensitive information detection. 
In~\cite{timmer2021sensitive}, we experiment with the fine-tuned Bidirectional Encoder Representations from Transformers (BERT)~\cite{devlin2018bert}. This method is graphically represented in Figure~\ref{fig:bert_visualisation}.  Experiments on the Monsanto trial data set show that the fine-tuned BERT model performs better than earlier approaches. 

\begin{figure}[tbp]
\centerline{\includegraphics[width=.8\linewidth]{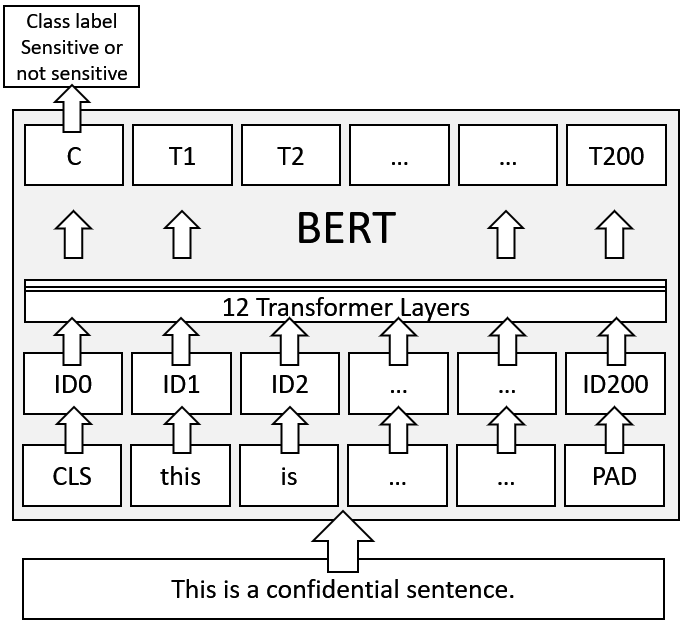}}
\caption{Visualisation of sensitive information detection with the BERT transformer.\label{fig:bert_visualisation}}
\end{figure}

Measuring the realism of a honeyfile is a complex task. There is a considerable research on realistic image~\cite{goodfellow2014generative, oord2016conditional} and text~\cite{cho2018towards, bakhtin2019real, bullock2019automated, karuna2018generating, johnson2018image} generation and, since Transformer models were released, on the perceived realism of generated text~\cite{clark2021all, brown2020language}. The realism of full documents has not been widely addressed, however. The combination of text, images, layout and format all contribute to the realism of a document, and the context in which it is observed is crucial. If, for example, a document repository search engine returns snapshots of the document cover or page on which the search results appear, the visual attributes will have a significant impact on the whether it is identified as a deception. 

We are currently conducting experiments testing perceived enticement and realism with crowd-sourced participants. 

\subsection{Simulating deceptive behaviours and interactions}
An important aspect of developing realistic deceptive environments is to simulate the various types of interactions seen on real enterprise networks, such as those on networking devices like routers, as well as employee interactions on direct messaging platforms like email and Slack/Teams. In particular, we want to develop generative models to simulate timestamped, directed communications between nodes on these networks. A natural choice for this task is Temporal Point Processes (TPPs), which are generative probabilistic models for event data with timestamps.

Depending on the type of network, communications may be uni-cast or multi-cast. A tool that would be especially handy to have in a cyber deception toolkit is an all-in-one model that has the ability to simulate any type of network communication, eg. uni-cast communications for a deceptive WiFi access point, or multi-cast communications for an email server honeypot. %We will discuss this problem and our work in this direction below.

\paragraph{Direct messaging}
Within organisations, email has evolved to be much more than just communications. Email platforms are typically integrated with task management tools which encourage the sharing of valuable company assets in the form of attachments in an email. This makes email servers an enticing target for attackers. The simulation of email networks can be split into two subtasks: 
\begin{enumerate}
    \item modeling when and with whom participants communicate, and
    \item generating the text to populate simulated conversation threads.
\end{enumerate} 
The first task can be tackled with TPPs, and the second task can be achieved with generative language models. %Email events can be represented as timestamped, directed hyperedges on the social network graph. 
A challenging aspect of the first subtask is that emails are \textit{multi-cast} communications, ie. they are directed from one sender to one or more recipients. If we consider a sender who is drafting an email, we can model their choice of whom to include as recipients as a multi-label classification problem. 

To the best of our knowledge, the interaction-partitioned topic model (IPTM) of Kim~\cite{kim2018thesis} is the only model in the literature that generates message events as well as their textual content. IPTM generates Multi-cast communications using the common multi-label classification approach, which is to model hyperedges as collections of independent edges, and then use binary classification independently on each edge incident to the sender node to decide whether or not to include that person in the recipient set. This modeling approach presents limitations in generating realistic events, namely it can lead to over or underestimation of event rates for communications with more than one recipient~\cite{chodrow2020configuration}, and can generate participant sets that were not seen in the training data. In addition to this, IPTM is only able to generate multi-cast communications, and therefore is not suitable for the all-in-one network communication simulator we seek to create. A further limitation of IPTM for application in cyber deception is that it does not generate human readable conversation text. Instead it uses a topic model to generate a set of word counts for each word in the vocabulary under consideration, which would immediately be identified as a deception by an intruder.

In~\cite{moore2021modelling}, we develop a direct messaging network simulation model for use in cyber deception. To increase the realism of generated deceptive communications, we introduce the LogNormMix-Net temporal point process, which learns edge weights for each hyperedge (as opposed to each pairwise edge as in IPTM~\cite{kim2018thesis}). This is achieved by modeling recipient selection as a multi-class classification problem, instead of multi-label binary classification in IPTM. This recipient selection approach also means that the LogNormMix-Net is able to simulate either uni-cast or multi-cast events, fulfilling the requirement for an all-in-one model that can simulate any type of network communication. 

We address the second subtask by applying a fine-tuned GPT-2 model to simulate an office email network where communication threads are coherent and stay on topic, and where each individual will have consistent themes to their communications, that are appropriate to their role within the simulated organisation. An example email and reply generated using the LogNormMix-Net together with our email thread generation model is presented in Figure \ref{fig:email-example}.

\begin{figure}[tbp]
\centerline{\includegraphics[width=8.5cm]{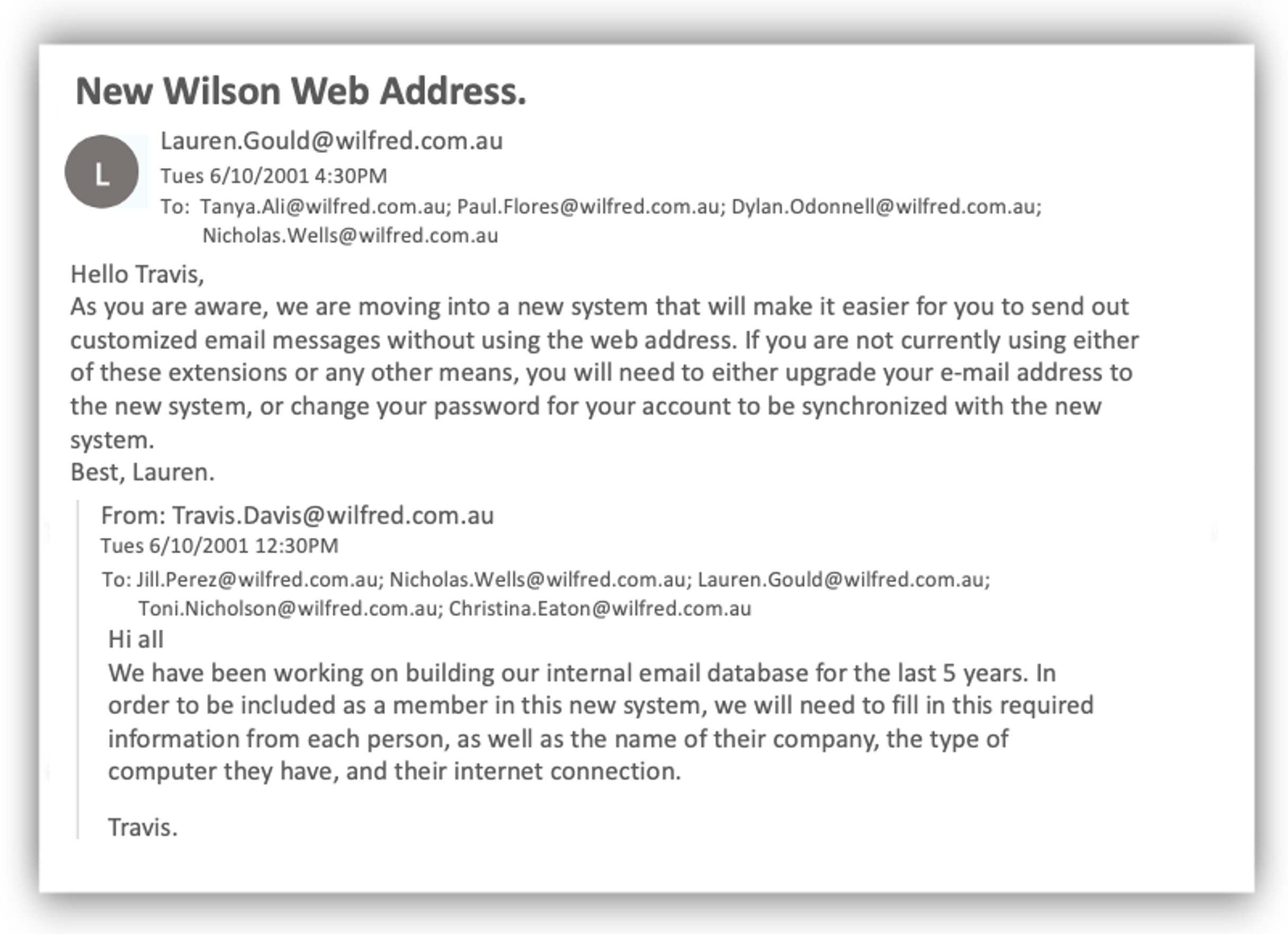}}
\caption{An example email thread generated from our direct messaging network simulation model.}
\label{fig:email-example}
\end{figure}

As a next step, we plan to utilize the LogNormMix-Net to generate other types of deception interactions, such as simulating WiFi access points.

\subsection{Ongoing and Future Work}
\label{sec:future}
We have a number of projects that are still in the early stages of development. 
\paragraph{Databases}
The generation of deceptive relational databases can be addressed with two tasks: firstly the generation of relational schemas, including table and column names, and secondly, populating the tables with conforming data (satisfying the schema design and constraints). To the best of our knowledge, there is no existing work that performs both of these tasks to generate completely novel databases. There is work in relational data augmentation~\cite{patki2016synthetic}, however these approaches are not designed to generate consistent and believable data from scratch.

%In addition to the contents of database tables, the structure of private datasets (eg. table and column names) may be considered sensitive, and as such it is desirable to be able to generate realistic relational schemas as well.
As a first step towards tackling database schema generation, we have extracted a dataset of relational schema~\cite{christopher2021schemadb}, as we could not find a suitably sized publicly available collection. Given a suitable schema training dataset, we think there are a couple of potentially viable training approaches. Relational structures naturally lend themselves to a graph representation as shown in Figure~\ref{fig:db_graph}, where edges link tables that are keyed. One potential approach is therefore to use contemporary graph generation models (such as those we use for code repository and wiki generation). These approaches are designed for homogeneous graph types though, whereas relational schemas are naturally heterogeneous structures, so we believe that they will be better modelled by heterogeneous graphs. 

Existing work in heterogeneous graph machine learning is primarily concerned with clustering, classification and edge detection, with successful generative approaches confined to specific domains (e.g. molecule generation)~\cite{wang2020survey,guo2020systematic,wu2019comprehensive,faez2021deep}. Therefore we see an opportunity in the development of an agnostic heterogeneous graph generator.

%An alternative possibility for schema generation is to take a language-based approach, instead of graph-based. Inspired by the ImageGPT work~\cite{chen2020generative}, we suggest that generating table content by fine-tuning a generative model like GPT-2 might be feasible. 

\begin{figure}[tbp]
\centerline{\includegraphics[scale=0.5]{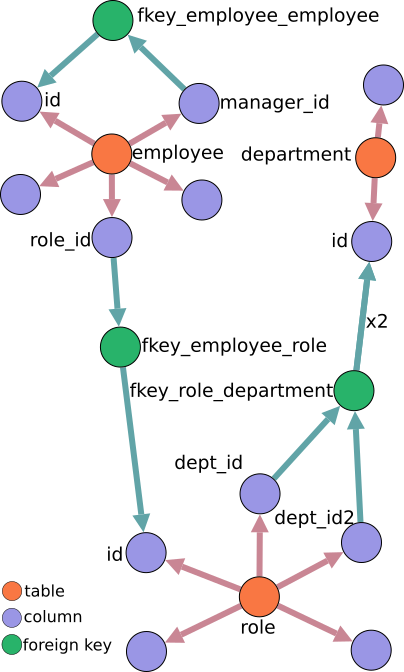}}
\caption{Graph representation of a relational schema. Some nodes left unlabelled for brevity}
\label{fig:db_graph}
\end{figure}

\paragraph{Interpretable generative models}
Control over content generation is increasingly important as the need for realism grows. Many generative approaches encode a compressed representation of the model domain called the latent space~\cite{kingma2019introduction}. Artefacts are generated by sampling from the latent space and {\em decoding} back into the model domain. This process does not in general, capture interpretable features in the latent space, and so does not allow for fine control over the characteristics of the generated artefacts. 

In this project, we study methods of disentangling that enable us to have this control over the generation. For example, consider the images shown in Figure~\ref{fig:disentangling}. This figure is an output from an extension of VAE/GAN model~\cite{srivastava2017veegan} whose latent variables are shape and colour. Each image is generated by sampling from latent shape and colour variables independently and decoding.  We note that changing the shape (variously colour) representation as we move left (variously top) to right (bottom) keeps the colour (shape) content the same, but changes the shoe shape (colour). In representation learning this property is known as a disentanglement; changes in a latent variable of a generative model explicitly changes a distinct aspect of the generated artefact~\cite{bengio2013representation}.

Our goal is to establish methods to bias generative models to represent features of interest to us in the latent space as latent variables and in a disentangled form. This allows us to alter features of the generated data independently to match a desired distribution. However, disentangled generative models cannot consistently be obtained using unsupervised learning~\cite{locatello2019challenging}, so we focus on semi-supervised methods to disentangle for interpretable features.

\begin{figure}[tbp]
\centerline{\includegraphics[width=\linewidth]{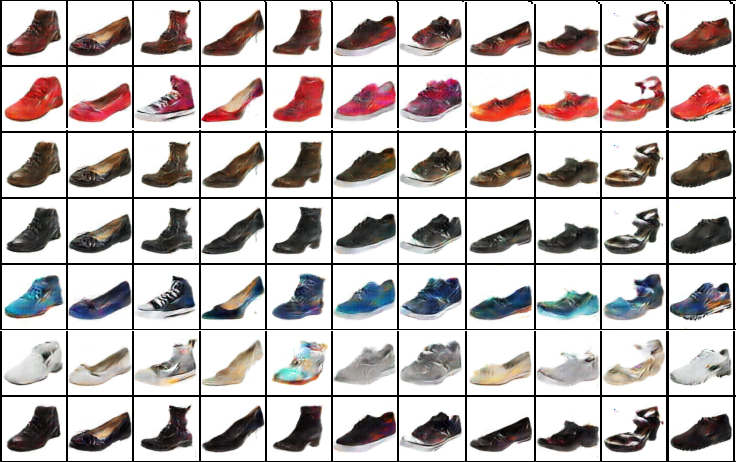}}
\caption{Disentangling colors (rows) and shapes (columns) on shoes data\cite{yu2014fine,yu2017semantic}}
\label{fig:disentangling}
\end{figure}

\paragraph{Graph generation}
Observing that graph representations are useful in contexts including file systems, database schema (see \S~\ref{sec:future}-a), participant users in a communications network  and wiki page connection, we're interested in indirect graph generation approaches. Recent work on using deep learning in graph generation can model attributed graphs~\cite{faez2021deep}. While these have performed well, there has been little research into allowing generation conditional on some input that informs the type or structure of the graph. This is of particular importance when the items in the training data are highly diverse, and the model usage would benefit from controlling the generated graph to some extent. To date, only one paper explores using a simple attribute vector for context in generation~\cite{yang2019conditional}. Similar to how an image has one or more natural text descriptions, graphs may have a textual description too, but no work to date incorporates long or complex attributes for conditioning. OpenAI's DALL-E~\cite{ramesh2021zero} demonstrates impressive text-to-image generation, with a user-supplied text input used to generate an matching image. We are investigating a similar approach for graphs, where a user-supplied text input is used to generate a matching graph.

In particular, generating temporal networks that represent a social network with communications between the (synthetic) participants presents an opportunity for exploring the benefits of text conditioning on generated graphs. The communication dynamics and propagation of topics within social networks is correlated with the topic and type of communication. Allowing a model to learn and depend on a given topic may improve realism significantly.

\paragraph{Availability of Datasets}
The simulation of rich and realistic environments requires models for a large range of artefacts and activities. One of the biggest limitations we have come across in achieving this goal is the scarcity of publicly available datasets for several of the artefacts or user behaviours we wish to generate. 

For some of these tasks, the data is not available for reasons of privacy and security - such as timestamped system logs of user actions. For other tasks, several individual examples can be found on the internet - e.g. database schemas, but tools must be created to locate and extract the individual examples, and then transform them into a standardised collection that is appropriate for ML tasks. 

There are therefore two open problems for datasets. The first is the release of new datasets that meet trustworthy and responsible AI requirements, including fairness and privacy. Examples of existing research works in this area include netflow datasets that are collected from real networks and then anonymised and released by the network owner~\cite{turcotte2018unified}. In the absence of access to real-world data sources, generating synthetic data from a cyber range may also be a possibility~\cite{moustafa2015milcom}. The second open problem is the development of tools that can be used to extract new datasets. Recent examples of extraction-based datasets include WIT: Wikipedia-based Image Text Dataset~\cite{srinivasan2021wit} and CC12M: Conceptual 12M~\cite{changpinyo2021conceptual}.

Some examples of our work in this direction includes:
\begin{itemize}
    \item Development of a sophisticated chart extraction tool to enable the extraction of a captioned chart dataset.
    \item Development of the \textsc{SchemaDB} dataset - a collection of relational schema in both \verb|MySQL| and heterogeneous graph form~\cite{christopher2021schemadb}.
    \item Development of a honeyfile corpus for use in experiments on measuring the enticement of honeyfiles~\cite{timmer2022tsm}.
\end{itemize}

\section{Discussions and concluding remarks}
In this work we have demonstrated various ways to use ML for generating deceptive content, and have highlighted some of the complexities and challenges. Generating realistic content and devising metrics to measure realism remains a challenge, as does finding datasets with which to train models. We recognise that content generation approaches that are privacy preserving, and which prevent language models like GPT from generating toxic content~\cite{solaiman2021process,abid2021persistent} are going to be increasingly important issues that we should pay attention to.

We have also discovered uses for generated content and behaviour outside of deception. 
Cyber range environments are a rapidly growing application domain for realistic simulations. Ranges have a host of important applications, including security testing and research, security education and capability development~\cite{european2020understanding}, all of which rely on content generation to make each environment distinct and convincing. These kinds of {\em digital twins} are increasingly used in a host of simulation activities~\cite{holmes2021digital}.

Another area that is evolving is Autonomous Cyber Operations (ACO). There are a number of ACO frameworks, much like cyber ranges, designed for training autonomous attacker and defender  agents~\cite{standen2021cyborg,molina2021network,microsoft2021cyberbattlesim,schwartz2019autonomous}. Since simulated ACO environments abstract away information that may be critical to an agent’s effectiveness, ACO agents can experience a `reality gap' between their performance on the simulation environments they are trained on compared to a real network environment~\cite{standen2021cyborg}. Making ACO environments as realistic as possible is one step to reduce this phenomenon. ACO environments are starting to incorporate features like benign user (gray agent) simulation~\cite{molina2021network} to make environments richer and more complex, but their behaviours are still quite limited, and several ACO training platform developers have emphasised the importance of, and opportunity for, improvement in the realism of their simulations\cite{standen2021cyborg,microsoft2021cyberbattlesim,schwartz2019autonomous}. 

The behaviour of an intruder faced with cyber deception is another area of considerable interest. It seems likely that awareness of
the possibility that any interaction with a system or artefact could reveal them causes greater caution and evaluation of the risk of each interaction. The behaviour and psychology 
of the adversary in the presence of cyber deception is increasingly a subject of study \cite{ferguson2018tularosa,ferguson2021decoy, ashenden2021design}. Even at this early stage, however,
it is fair to say that, from the defender's perspective, anything that makes intrusion and theft more 
costly for the perpetrator should be encouraged. 

In closing we should note that, despite extolling the virtues and benefits of deception throughout this paper, we are not advocating it 
as an alternative to more common security measures. Rather, deception is a complement to existing, perimeter
focused security. It acknowledges the risk that, even with the best security, breaches will happen, and 
provides an additional layer of protection through early detection, adversary intelligence and disruption.  

\section{Acknowledgements}
The work has been supported by the Cyber Security Research Centre Limited whose activities are partially funded by the Australian Government’s Cooperative Research Centres Programme. The authors would like to thank the students who participated in the project:
Ricard Grace, Alexander Bunn and Adam Green, as well as the CSCRC vacation students Renee Selvey, Duy Khuu and David Liu.

\newpage
\bibliographystyle{IEEEtran}
\bibliography{deception}

\begin{thebibliography}{10}
\providecommand{\url}[1]{#1}
\csname url@samestyle\endcsname
\providecommand{\newblock}{\relax}
\providecommand{\bibinfo}[2]{#2}
\providecommand{\BIBentrySTDinterwordspacing}{\spaceskip=0pt\relax}
\providecommand{\BIBentryALTinterwordstretchfactor}{4}
\providecommand{\BIBentryALTinterwordspacing}{\spaceskip=\fontdimen2\font plus
\BIBentryALTinterwordstretchfactor\fontdimen3\font minus
  \fontdimen4\font\relax}
\providecommand{\BIBforeignlanguage}[2]{{%
\expandafter\ifx\csname l@#1\endcsname\relax
\typeout{** WARNING: IEEEtran.bst: No hyphenation pattern has been}%
\typeout{** loaded for the language `#1'. Using the pattern for}%
\typeout{** the default language instead.}%
\else
\language=\csname l@#1\endcsname
\fi
#2}}
\providecommand{\BIBdecl}{\relax}
\BIBdecl

\bibitem{gartner_spending_2021}
\BIBentryALTinterwordspacing
Gartner. (2017) Gartner forecasts worldwide security and risk management
  spending to exceed \$150 billion in 2021. [Online]. Available:
  \url{https://www.gartner.com/en/newsroom/press-releases/2021-05-17-gartner-forecasts-worldwide-security-and-risk-managem}
\BIBentrySTDinterwordspacing

\bibitem{weiss2015target}
N.~E. Weiss and R.~S. Miller, ``The target and other financial data breaches:
  Frequently asked questions,'' in \emph{Congressional Research Service,
  Prepared for Members and Committees of Congress February}, vol.~4, 2015, p.
  2015.

\bibitem{targetpressrelease2013}
\BIBentryALTinterwordspacing
{Target Corporation}. (2013) Target confirms unauthorized access to payment
  card data in u.s. stores. [Online]. Available:
  \url{https://corporate.target.com/press/releases/2013/12/target-confirms-unauthorized-access-to-payment-car}
\BIBentrySTDinterwordspacing

\bibitem{krebs2014target}
\BIBentryALTinterwordspacing
B.~Krebs. (2014) The target breach, by the numbers. [Online]. Available:
  \url{https://krebsonsecurity.com/2014/05/the-target-breach-by-the-numbers/}
\BIBentrySTDinterwordspacing

\bibitem{reuters2017target}
\BIBentryALTinterwordspacing
Reuters. (2017) Target settles 2013 hacked customer data breach for {\$ 18.5}
  million. [Online]. Available:
  \url{www.nbcnews.com/business/business-news/target-settles-2013-hacked-customer-data-breach-18-5-million-n764031}
\BIBentrySTDinterwordspacing

\bibitem{harrell2017synergistic}
M.~N. Harrell, ``Synergistic security: A work system case study of the target
  breach,'' \emph{Journal of Cybersecurity Education, Research and Practice},
  vol. 2017, no.~2, p.~4, 2017.

\bibitem{plachkinova2018security}
M.~Plachkinova and C.~Maurer, ``Security breach at target,'' \emph{Journal of
  Information Systems Education}, vol.~29, no.~1, pp. 11--20, 2018.

\bibitem{schwartz2014target}
\BIBentryALTinterwordspacing
M.~J. Schwartz. (2014) Target ignored data breach alarms. [Online]. Available:
  \url{https://www.darkreading.com/attacks-and-breaches/target-ignored-data-breach-alarms/d/d-id/1127712}
\BIBentrySTDinterwordspacing

\bibitem{ponemon2021}
{Ponemon Institute}, ``2021 cost of a data breach study: Global overview,''
  June 2021.

\bibitem{spitzner2003-1-honeypots}
L.~Spitzner, ``Honeypots: Catching the insider threat,'' in \emph{19th Annu.
  Computer Security Applicat. Conf., Proc.}\hskip 1em plus 0.5em minus
  0.4em\relax IEEE, 2003, pp. 170--179.

\bibitem{fraunholz2018defending}
D.~Fraunholz and H.~D. Schotten, ``Defending web servers with feints,
  distraction and obfuscation,'' in \emph{Proc. Int. Conf. Computing,
  Networking and Communications}.\hskip 1em plus 0.5em minus 0.4em\relax IEEE,
  2018, pp. 21--25.

\bibitem{stevens2016cheats}
M.~Stevens, \emph{Cheats and deceits: how animals and plants exploit and
  mislead}.\hskip 1em plus 0.5em minus 0.4em\relax Oxford University Press,
  2016.

\bibitem{smith2014overlord}
T.~J. Smith, ``Overlord/bodyguard: Intelligence failure through adversary
  deception,'' \emph{International Journal of Intelligence and
  CounterIntelligence}, vol.~27, no.~3, pp. 550--568, 2014.

\bibitem{bell2003toward}
J.~B. Bell, ``Toward a theory of deception,'' \emph{Int. J. of Intell. and
  CounterIntell.}, vol.~16, no.~2, pp. 244--279, 2003.

\bibitem{whaley1982toward}
B.~Whaley, ``Toward a general theory of deception,'' \emph{J. of Strategic
  Stud.}, vol.~5, no.~1, pp. 178--192, 1982.

\bibitem{stoll1988stalking}
C.~Stoll, ``Stalking the wily hacker,'' \emph{Communications of the ACM},
  vol.~31, no.~5, pp. 484--497, 1988.

\bibitem{stoll2005cuckoo}
------, \emph{The Cuckoo's Egg: Tracking a Spy through the Maze of Computer
  Espionage}.\hskip 1em plus 0.5em minus 0.4em\relax Simon and Schuster, 2005.

\bibitem{yuill2004honeyfiles}
J.~Yuill, M.~Zappe, D.~Denning, and F.~Feer, ``Honeyfiles: deceptive files for
  intrusion detection,'' in \emph{Proc. the 5th Annu. SMC Information Assurance
  Workshop}.\hskip 1em plus 0.5em minus 0.4em\relax IEEE, 2004, pp. 116--122.

\bibitem{voris2012lost}
J.~Voris, N.~Boggs, and S.~J. Stolfo, ``Lost in translation: Improving decoy
  documents via automated translation,'' in \emph{Security and Privacy
  Workshops, IEEE Symp. on}, 2012, pp. 129--133.

\bibitem{voris2013bait}
J.~Voris, J.~Jermyn, A.~D. Keromytis, and S.~J. Stolfo, ``Bait and snitch:
  Defending computer systems with decoys,'' in \emph{Proc. the cyber
  infrastructure protection Conf., Strategic Stud. Institute, September}, 2013.

\bibitem{voris2015fox}
J.~Voris, J.~Jermyn, N.~Boggs, and S.~Stolfo, ``Fox in the trap: thwarting
  masqueraders via automated decoy document deployment,'' in \emph{Proc. the
  8th European Workshop on System Security}.\hskip 1em plus 0.5em minus
  0.4em\relax ACM, 2015, p.~3.

\bibitem{bowen2009baiting}
B.~M. Bowen, S.~Hershkop, A.~D. Keromytis, and S.~J. Stolfo, ``Baiting inside
  attackers using decoy documents.'' in \emph{SecureComm}, vol.~19.\hskip 1em
  plus 0.5em minus 0.4em\relax Springer, 2009, pp. 51--70.

\bibitem{salem2011decoy}
M.~B. Salem and S.~J. Stolfo, ``Decoy document deployment for effective
  masquerade attack detection,'' in \emph{Int. Conf. Detection of Intrusions
  and Malware, and Vulnerability Assessment}.\hskip 1em plus 0.5em minus
  0.4em\relax Springer, 2011, pp. 35--54.

\bibitem{whitham2017automating}
B.~Whitham, ``Automating the generation of enticing text content for
  high-interaction honeyfiles,'' in \emph{Proc. the 50th Hawaii Int. Conf.
  System Sciences}, 2017.

\bibitem{bercovitch2011honeygen}
M.~Bercovitch, M.~Renford, L.~Hasson, A.~Shabtai, L.~Rokach, and Y.~Elovici,
  ``Honeygen: An automated honeytokens generator,'' in \emph{Proceedings of
  2011 IEEE International Conference on Intelligence and Security
  Informatics}.\hskip 1em plus 0.5em minus 0.4em\relax IEEE, 2011, pp.
  131--136.

\bibitem{abay2019using}
N.~C. Abay, C.~G. Akcora, Y.~Zhou, M.~Kantarcioglu, and B.~Thuraisingham,
  ``Using deep learning to generate relational honeydata,'' in \emph{Autonomous
  Cyber Deception}.\hskip 1em plus 0.5em minus 0.4em\relax Springer, 2019, pp.
  3--19.

\bibitem{radford2018improving}
A.~Radford, K.~Narasimhan, T.~Salimans, and I.~Sutskever, ``Improving language
  understanding by generative pre-training,'' 2018.

\bibitem{graves2013generating}
A.~Graves, ``Generating sequences with recurrent neural networks,'' \emph{arXiv
  preprint arXiv:1308.0850}, 2013.

\bibitem{shchur2021review}
\BIBentryALTinterwordspacing
O.~Shchur, A.~C. T{\"{u}}rkmen, T.~Januschowski, and S.~G{\"{u}}nnemann,
  ``Neural temporal point processes: {A} review,'' \emph{CoRR}, vol.
  abs/2104.03528, 2021. [Online]. Available:
  \url{https://arxiv.org/abs/2104.03528}
\BIBentrySTDinterwordspacing

\bibitem{liao2019efficient}
R.~Liao, Y.~Li, Y.~Song, S.~Wang, W.~Hamilton, D.~K. Duvenaud, R.~Urtasun, and
  R.~Zemel, ``Efficient graph generation with graph recurrent attention
  networks,'' in \emph{Advances in Neural Information Processing Systems},
  2019, pp. 4257--4267.

\bibitem{nguyen2021honeycode}
D.~Nguyen, D.~Liebowitz, S.~Nepal, and S.~Kanhere, ``Honeycode: Automating
  deceptive software repositories with deep generative models,'' in \emph{Proc.
  the 54th Hawaii Int. Conf. Syst. Sci.}, 2021, p. 6945.

\bibitem{vaswani2017attention}
A.~Vaswani, N.~Shazeer, N.~Parmar, J.~Uszkoreit, L.~Jones, A.~N. Gomez,
  {\L}.~Kaiser, and I.~Polosukhin, ``Attention is all you need,'' in
  \emph{Advances in neural information processing systems}, 2017, pp.
  5998--6008.

\bibitem{longland2020wikigen}
\BIBentryALTinterwordspacing
M.~Longland, ``Generating fake websites: Wikigen,'' Australian National
  University, Tech. Rep., 2020. [Online]. Available:
  \url{https://github.com/longland-m/wikigen}
\BIBentrySTDinterwordspacing

\bibitem{arora2017asimple}
S.~Arora, Y.~Liang, and T.~Ma, ``A simple but tough-to-beat baseline for
  sentence embeddings,'' in \emph{International conference on learning
  representations}, 2017.

\bibitem{li2019layoutgan}
J.~Li, J.~Yang, A.~Hertzmann, J.~Zhang, and T.~Xu, ``Layoutgan: Generating
  graphic layouts with wireframe discriminators,'' \emph{arXiv preprint
  arXiv:1901.06767}, 2019.

\bibitem{jyothi2019layoutvae}
A.~A. Jyothi, T.~Durand, J.~He, L.~Sigal, and G.~Mori, ``Layoutvae: Stochastic
  scene layout generation from a label set,'' in \emph{Proceedings of the
  IEEE/CVF International Conference on Computer Vision}, 2019, pp. 9895--9904.

\bibitem{lee2020neural}
H.-Y. Lee, L.~Jiang, I.~Essa, P.~B. Le, H.~Gong, M.-H. Yang, and W.~Yang,
  ``Neural design network: Graphic layout generation with constraints,'' in
  \emph{Computer Vision--ECCV 2020: 16th European Conference, Glasgow, UK,
  August 23--28, 2020, Proceedings, Part III 16}.\hskip 1em plus 0.5em minus
  0.4em\relax Springer, 2020, pp. 491--506.

\bibitem{kikuchi2021constrained}
K.~Kikuchi, E.~Simo-Serra, M.~Otani, and K.~Yamaguchi, ``Constrained graphic
  layout generation via latent optimization,'' in \emph{Proceedings of the 29th
  ACM International Conference on Multimedia}, 2021, pp. 88--96.

\bibitem{nguyen2021diverse}
D.~D. Nguyen, S.~Nepal, and S.~S. Kanhere, ``Diverse multimedia layout
  generation with multi choice learning,'' in \emph{Proceedings of the 29th ACM
  International Conference on Multimedia}, 2021, pp. 218--226.

\bibitem{guzman2012multiple}
A.~Guzman-Rivera, D.~Batra, and P.~Kohli, ``Multiple choice learning: Learning
  to produce multiple structured outputs.'' in \emph{NIPS}, vol.~1,
  no.~2.\hskip 1em plus 0.5em minus 0.4em\relax Citeseer, 2012, p.~3.

\bibitem{timmer2022tsm}
R.~Timmer, D.~Liebowitz, S.~Nepal, and S.~Kanhere, ``Tsm: Measuring the
  enticement of honeyfiles with natural language processing,'' in \emph{Proc.
  the 55th Hawaii Int. Conf. Syst. Sci.}, 2022.

\bibitem{cho2008detectingprivacy}
R.~Chow, P.~Golle, and J.~Staddon, ``Detecting privacy leaks using corpus-based
  association rules,'' in \emph{Proceedings of the 14th ACM SIGKDD
  international conference on Knowledge discovery and data mining}, 2008, pp.
  893--901.

\bibitem{sanchez2016c}
D.~S{\'a}nchez and M.~Batet, ``C-sanitized: A privacy model for document
  redaction and sanitization,'' \emph{Journal of the Association for
  Information Science and Technology}, vol.~67, no.~1, pp. 148--163, 2016.

\bibitem{neerbek2018detecting}
J.~Neerbek, I.~Assent, and P.~Dolog, ``Detecting complex sensitive information
  via phrase structure in recursive neural networks,'' in \emph{Pacific-Asia
  Conf. Knowledge Discovery and Data Mining}.\hskip 1em plus 0.5em minus
  0.4em\relax Springer, 2018, pp. 373--385.

\bibitem{timmer2021sensitive}
R.~Timmer, D.~Liebowitz, S.~Nepal, and S.~Kanhere, ``Can pre-trained
  transformers be used in detecting complex sensitive sentences? - a monsanto
  case study,'' in \emph{Proc. the 3rd IEEE International Conference on Trust,
  Privacy and Security in Intelligent Systems, and Applications}, 2021.

\bibitem{devlin2018bert}
J.~Devlin, M.-W. Chang, K.~Lee, and K.~Toutanova, ``Bert: Pre-training of deep
  bidirectional transformers for language understanding,'' \emph{arXiv preprint
  arXiv:1810.04805}, 2018.

\bibitem{goodfellow2014generative}
I.~Goodfellow, J.~Pouget-Abadie, M.~Mirza, B.~Xu, D.~Warde-Farley, S.~Ozair,
  A.~Courville, and Y.~Bengio, ``Generative adversarial nets,'' in
  \emph{Advances in neural information processing systems}, 2014, pp.
  2672--2680.

\bibitem{oord2016conditional}
A.~v.~d. Oord, N.~Kalchbrenner, O.~Vinyals, L.~Espeholt, A.~Graves, and
  K.~Kavukcuoglu, ``Conditional image generation with pixelcnn decoders,''
  \emph{arXiv preprint arXiv:1606.05328}, 2016.

\bibitem{cho2018towards}
W.~S. Cho, P.~Zhang, Y.~Zhang, X.~Li, M.~Galley, C.~Brockett, M.~Wang, and
  J.~Gao, ``Towards coherent and cohesive long-form text generation,''
  \emph{arXiv preprint arXiv:1811.00511}, 2018.

\bibitem{bakhtin2019real}
A.~Bakhtin, S.~Gross, M.~Ott, Y.~Deng, M.~Ranzato, and A.~Szlam, ``Real or
  fake? learning to discriminate machine from human generated text,''
  \emph{arXiv preprint arXiv:1906.03351}, 2019.

\bibitem{bullock2019automated}
J.~Bullock and M.~Luengo-Oroz, ``Automated speech generation from un general
  assembly statements: Mapping risks in ai generated texts,'' \emph{arXiv
  preprint arXiv:1906.01946}, 2019.

\bibitem{karuna2018generating}
P.~Karuna, H.~Purohit, R.~Ganesan, and S.~Jajodia, ``Generating hard to
  comprehend fake documents for defensive cyber deception,'' \emph{IEEE Intell.
  Syst.}, vol.~33, no.~5, pp. 16--25, 2018.

\bibitem{johnson2018image}
J.~Johnson, A.~Gupta, and L.~Fei-Fei, ``Image generation from scene graphs,''
  in \emph{Proceedings of the IEEE conference on computer vision and pattern
  recognition}, 2018, pp. 1219--1228.

\bibitem{clark2021all}
E.~Clark, T.~August, S.~Serrano, N.~Haduong, S.~Gururangan, and N.~A. Smith,
  ``All that's' human'is not gold: Evaluating human evaluation of generated
  text,'' \emph{arXiv preprint arXiv:2107.00061}, 2021.

\bibitem{brown2020language}
T.~B. Brown, B.~Mann, N.~Ryder, M.~Subbiah, J.~Kaplan, P.~Dhariwal,
  A.~Neelakantan, P.~Shyam, G.~Sastry, A.~Askell \emph{et~al.}, ``Language
  models are few-shot learners,'' \emph{arXiv preprint arXiv:2005.14165}, 2020.

\bibitem{kim2018thesis}
B.~Kim, ``Latent modeling of dynamic social networks,'' Ph.D. dissertation, The
  Pennsylvania State University, 8 2018.

\bibitem{chodrow2020configuration}
P.~S. Chodrow, ``Configuration models of random hypergraphs,'' \emph{Journal of
  Complex Networks}, vol.~8, no.~3, p. cnaa018, 2020.

\bibitem{moore2021modelling}
K.~Moore, C.~J. Christopher, D.~Liebowitz, S.~Nepal, and R.~Selvey, ``Modelling
  direct messaging networks with multiple recipients for cyber deception,''
  \emph{arXiv preprint arXiv:2111.11932}, 2021.

\bibitem{patki2016synthetic}
N.~Patki, R.~Wedge, and K.~Veeramachaneni, ``The synthetic data vault,'' in
  \emph{2016 IEEE International Conference on Data Science and Advanced
  Analytics (DSAA)}.\hskip 1em plus 0.5em minus 0.4em\relax IEEE, 2016, pp.
  399--410.

\bibitem{christopher2021schemadb}
C.~J. Christopher, K.~Moore, and D.~Liebowitz, ``\textsc{SchemaDB}: Structures
  in relational datasets,'' \emph{arXiv preprint arXiv:2111.12835}, 2021.

\bibitem{wang2020survey}
X.~Wang, D.~Bo, C.~Shi, S.~Fan, Y.~Ye, and P.~S. Yu, ``A survey on
  heterogeneous graph embedding: methods, techniques, applications and
  sources,'' \emph{arXiv preprint arXiv:2011.14867}, 2020.

\bibitem{guo2020systematic}
X.~Guo and L.~Zhao, ``A systematic survey on deep generative models for graph
  generation,'' \emph{arXiv preprint arXiv:2007.06686}, 2020.

\bibitem{wu2019comprehensive}
Z.~Wu, S.~Pan, F.~Chen, G.~Long, C.~Zhang, and P.~S. Yu, ``A comprehensive
  survey on graph neural networks,'' \emph{arXiv preprint arXiv:1901.00596},
  2019.

\bibitem{faez2021deep}
F.~Faez, Y.~Ommi, M.~S. Baghshah, and H.~R. Rabiee, ``Deep graph generators: A
  survey,'' \emph{IEEE Access}, vol.~9, pp. 106\,675--106\,702, 2021.

\bibitem{kingma2019introduction}
D.~P. Kingma and M.~Welling, ``An introduction to variational autoencoders,''
  \emph{arXiv preprint arXiv:1906.02691}, 2019.

\bibitem{srivastava2017veegan}
A.~Srivastava, L.~Valkov, C.~Russell, M.~U. Gutmann, and C.~Sutton, ``Veegan:
  Reducing mode collapse in gans using implicit variational learning,'' in
  \emph{Proceedings of the 31st International Conference on Neural Information
  Processing Systems}, 2017, pp. 3310--3320.

\bibitem{bengio2013representation}
Y.~Bengio, A.~Courville, and P.~Vincent, ``Representation learning: A review
  and new perspectives,'' \emph{IEEE transactions on pattern analysis and
  machine intelligence}, vol.~35, no.~8, pp. 1798--1828, 2013.

\bibitem{locatello2019challenging}
F.~Locatello, S.~Bauer, M.~Lucic, G.~Raetsch, S.~Gelly, B.~Sch{\"o}lkopf, and
  O.~Bachem, ``Challenging common assumptions in the unsupervised learning of
  disentangled representations,'' in \emph{Int. Conf. machine learning}.\hskip
  1em plus 0.5em minus 0.4em\relax PMLR, 2019, pp. 4114--4124.

\bibitem{yu2014fine}
A.~Yu and K.~Grauman, ``Fine-grained visual comparisons with local learning,''
  in \emph{Proceedings of the IEEE Conference on Computer Vision and Pattern
  Recognition}, 2014, pp. 192--199.

\bibitem{yu2017semantic}
------, ``Semantic jitter: Dense supervision for visual comparisons via
  synthetic images,'' in \emph{Proceedings of the IEEE International Conference
  on Computer Vision}, 2017, pp. 5570--5579.

\bibitem{yang2019conditional}
C.~Yang, P.~Zhuang, W.~Shi, A.~Luu, and P.~Li, ``Conditional structure
  generation through graph variational generative adversarial nets.'' in
  \emph{NeurIPS}, 2019, pp. 1338--1349.

\bibitem{ramesh2021zero}
A.~Ramesh, M.~Pavlov, G.~Goh, S.~Gray, C.~Voss, A.~Radford, M.~Chen, and
  I.~Sutskever, ``Zero-shot text-to-image generation,'' \emph{arXiv preprint
  arXiv:2102.12092}, 2021.

\bibitem{turcotte2018unified}
\BIBentryALTinterwordspacing
M.~J.~M. Turcotte, A.~D. Kent, and C.~Hash, \emph{Unified Host and Network Data
  Set}.\hskip 1em plus 0.5em minus 0.4em\relax World Scientific, 11 2018, ch.
  Chapter 1, pp. 1--22. [Online]. Available:
  \url{https://www.worldscientific.com/doi/abs/10.1142/9781786345646_001}
\BIBentrySTDinterwordspacing

\bibitem{moustafa2015milcom}
N.~Moustafa and J.~Slay, ``Unsw-nb15: a comprehensive data set for network
  intrusion detection systems (unsw-nb15 network data set),'' in \emph{2015
  Military Communications and Information Systems Conference (MilCIS)}, 2015,
  pp. 1--6.

\bibitem{srinivasan2021wit}
K.~Srinivasan, K.~Raman, J.~Chen, M.~Bendersky, and M.~Najork, ``Wit:
  Wikipedia-based image text dataset for multimodal multilingual machine
  learning,'' 2021.

\bibitem{changpinyo2021conceptual}
S.~Changpinyo, P.~Sharma, N.~Ding, and R.~Soricut, ``Conceptual 12m: Pushing
  web-scale image-text pre-training to recognize long-tail visual concepts,''
  2021.

\bibitem{solaiman2021process}
I.~Solaiman and C.~Dennison, ``Process for adapting language models to society
  (palms) with values-targeted datasets,'' 2021.

\bibitem{abid2021persistent}
A.~Abid, M.~Farooqi, and J.~Zou, ``Persistent anti-muslim bias in large
  language models,'' \emph{arXiv preprint arXiv:2101.05783}, 2021.

\bibitem{european2020understanding}
\BIBentryALTinterwordspacing
E.~C. S.~O. (ECSO). (2020) Understanding cyber ranges: From hype to reality.
  ECS Brussels, Belgium. [Online]. Available:
  \url{https://www.ecs-org.eu/documents/uploads/understanding-cyber-ranges-from-hype-to-reality.pdf}
\BIBentrySTDinterwordspacing

\bibitem{holmes2021digital}
D.~Holmes, M.~Papathanasaki, L.~Maglaras, M.~A. Ferrag, S.~Nepal, and
  H.~Janicke, ``Digital twins and cyber security--solution or challenge?'' in
  \emph{2021 6th South-East Europe Design Automation, Computer Engineering,
  Computer Networks and Social Media Conference (SEEDA-CECNSM)}.\hskip 1em plus
  0.5em minus 0.4em\relax IEEE, 2021, pp. 1--8.

\bibitem{standen2021cyborg}
M.~Standen, M.~Lucas, D.~Bowman, T.~J. Richer, J.~Kim, and D.~Marriott,
  ``Cyborg: A gym for the development of autonomous cyber agents,'' \emph{arXiv
  preprint arXiv:2108.09118}, 2021.

\bibitem{molina2021network}
A.~Molina-Markham, C.~Miniter, B.~Powell, and A.~Ridley, ``Network environment
  design for autonomous cyberdefense,'' \emph{arXiv preprint arXiv:2103.07583},
  2021.

\bibitem{microsoft2021cyberbattlesim}
M.~D.~R. Team \emph{et~al.}, ``Cyberbattlesim,'' 2021.

\bibitem{schwartz2019autonomous}
J.~Schwartz and H.~Kurniawati, ``Autonomous penetration testing using
  reinforcement learning,'' \emph{arXiv preprint arXiv:1905.05965}, 2019.

\bibitem{ferguson2018tularosa}
K.~Ferguson-Walter, T.~Shade, A.~Rogers, M.~C.~S. Trumbo, K.~S. Nauer, K.~M.
  Divis, A.~Jones, A.~Combs, and R.~G. Abbott, ``The tularosa study: An
  experimental design and implementation to quantify the effectiveness of cyber
  deception.'' Sandia National Lab.(SNL-NM), Albuquerque, NM (United States),
  Tech. Rep., 2018.

\bibitem{ferguson2021decoy}
K.~J. Ferguson-Walter, M.~M. Major, C.~K. Johnson, and D.~H. Muhleman,
  ``Examining the efficacy of decoy-based and psychological cyber deception,''
  in \emph{30th $\{$USENIX$\}$ Security Symposium ($\{$USENIX$\}$ Security
  21)}, 2021.

\bibitem{ashenden2021design}
D.~Ashenden, R.~Black, I.~Reid, and S.~Henderson, ``Design thinking for cyber
  deception,'' in \emph{Proceedings of the 54th Hawaii International Conference
  on System Sciences}, 2021, p. 1958.

\end{thebibliography}
\end{document}